\documentclass[aps,prl,showkeys,showpacs,amssymb,cite,
amsfonts,epsf,preprintnumbers,nofootinbib,superscriptaddress]{revtex4}

\usepackage[dvips]{graphicx}
\usepackage{bm,latexsym,amsmath,amssymb,amsfonts,color}

\newcommand\be{\begin{equation}}
\newcommand\ee{\end{equation}}
\newcommand\beq{\begin{eqnarray}}
\newcommand\eeq{\end{eqnarray}}

\newcommand{\nn}{\nonumber}

\newcommand{\cV}{\mathcal{V}}

\begin{document}

\title{A new synthesis of matter and gravity: A nongravitating scalar field}

\author{Inyong Cho}
\email{iycho@seoultech.ac.kr}
\affiliation{Institute of Convergence Fundamental Studies \& School of Liberal Arts,
Seoul National University of Science and Technology, Seoul 139-743, Korea}
\author{Hyeong-Chan Kim}
\email{hckim@ut.ac.kr}
\affiliation{School of Liberal Arts and Sciences, Korea National University of Transportation, Chungju 380-702, Korea}

\begin{abstract}
We present a new manifestation of the nonlinearity of the gravity-matter interactions.
We show explicitly that there exists a nongravitating dynamical scalar-field solution in Eddington-inspired Born-Infeld gravity.
This kind of solution has not been found in previous literatures based on general relativity, or other modified-gravity theories.
The nongravitating solution obtained in this work is unstable to perturbations.
\end{abstract}
\maketitle

The matter interacts with gravity closely.
In this interaction, the mutual role of the matter and gravity can be summarized by the word of J. A. Wheeler, ``Spacetime tells matter how to move; matter tells spacetime how to curve."
However, looking closely into the solution space of general relativity (GR), one may find that
the role played by gravity is not exactly the same as that played by the matter as a generator of gravity especially in the vacuum sector.
In GR, the nonlinearity of the gravity allows curved spacetimes in the absence of matter such as the gravitational soliton, gravitational instanton, Bianchi solutions, etc.
These solutions play crucial role in understanding the Einstein gravity.
Contrary to that, GR does not allow matter dynamics in the absence of gravity.
Several generalized gravity theories appeared in the literature such as the scalar-tensor theory, the $f(R)$ gravity,  the massive gravity, and the Gauss-Bonnet gravity.
However until now, no gravity theory is known to have the nonlinear interaction allowing a dynamical-matter field without developing gravity.
Therefore, it is worthwhile to investigate nongravitating matter distributions (nGMD) via nonlinear interactions between the matter and gravity in newly suggested gravity theories.

In gravity theories modified from the Einstein-Hilbert action by adding higher-curvature terms, one can easily notice that the nGMD may not exist.
The equation of motion for the action will take the form,
$$
\sum_{n=0} c_n (\mathcal{R}^n)_{\mu\nu} = T_{\mu\nu},
$$
where $\mathcal{R}^n$ represents the every $n^{\rm th}$-order polynomial of curvatures. Let us consider the regular situations when the curvatures vanish.
In the case of $c_0=0$, one may easily notice that the zero curvature simply leads to zero stress tensor $T_{\mu\nu}=0$.
If $c_0\neq 0$, the stress tensor plays the role of the cosmological constant, which implies the matter field is homogeneous and nondynamical.
Therefore, we may conclude that the gravity theory allowing nGMD should be based on a different starting point rather than simply including higher curvature terms.

In this letter, we show that
the Eddington-inspired Born-Infeld (EiBI) theory, which was suggested as an alternative theory of gravity recently~\cite{Banados:2010ix}, allows nGMD.
The EiBI action is given by
\begin{eqnarray}\label{maction}
S_{{\rm EiBI}}=\frac{1}{\kappa}\int
d^4x\Big[~\sqrt{-|g_{\mu\nu}+\kappa
R_{\mu\nu}(\Gamma)|}-\lambda\sqrt{-|g_{\mu\nu}|}~\Big]+S_M(g,\Phi),
\end{eqnarray}
where
$|{\cal G}_{\mu\nu}|$ denotes the determinant of ${\cal G}_{\mu\nu}$,
$\lambda$  is a dimensionless parameter
related with the cosmological constant by $\Lambda = (\lambda -1)/\kappa$,
and we set $8\pi G=1$.
In this theory the metric $g_{\mu\nu}$ and the connection $\Gamma_{\mu\nu}^{\rho}$
are treated as independent fields (Palatini formalism).
The Ricci tensor $R_{\mu\nu}(\Gamma)$ is evaluated solely by the connection,
and the matter filed $\Phi$ is coupled only to the gravitational field $g_{\mu\nu}$.
The merits of this theory are that it requires only one more theory parameter $\kappa$,
and that it is equivalent to the theory of GR in vacuum.

In Refs.~\cite{Banados:2010ix,Cho:2012vg},
the evolution of the Universe driven by barotropic fluid
had been inverstigated in EiBI gravity.
For the equation-of-state parameter, $w\equiv P/\rho > 0$,
the Universe starts from a nonsingular initial state of a finite size for $\kappa>0$.
More interestingly, the initial state of the Universe driven by pressureless dust ($w=0$)
approaches the de Sitter state with the effective cosmological constant
$\Lambda_{\rm eff} = 8/\kappa$~\cite{Cho:2012vg}.
Subsequent works in EiBI gravity have been performed on the subject of
the cosmological and astrophysical constraints on the EiBI theory~\cite{DeFelice:2012hq,Avelino:2012ge},
the constraint on the value of $\kappa$ by using the solar model~\cite{Casanellas:2011kf},
the tensor perturbation \cite{EscamillaRivera:2012vz},
bouncing cosmology \cite{Avelino:2012ue},
the five dimensional brane model \cite{Liu:2012rc},
the effective stress tensor and energy conditions \cite{Delsate:2012ky},
cosmology with scalar fields~\cite{Scargill:2012kg},
the instability of compact stars~\cite{Sham:2012qi},
the surface singularity of the compact star~\cite{Pani:2012qd},
etc.

The equations of motion are obtained by varying the action \eqref{maction}
with respect to the fields $g_{\mu\nu}$ and $\Gamma_{\mu\nu}^{\rho}$ respectively,
\be\label{eom1}
\frac{\sqrt{-|q|}}{\sqrt{-|g|}}~q^{\mu\nu}
=\lambda g^{\mu\nu} -\kappa T^{\mu\nu},
\ee
and
\be\label{eom2}
q_{\mu\nu} = g_{\mu\nu}+\kappa R_{\mu\nu},
\ee
where $q_{\mu\nu}$ is the auxiliary metric by which
the connection $\Gamma_{\mu\nu}^{\rho}$ is defined,
and $q^{\mu\nu}$ is the matrix inverse of $q_{\mu\nu}$.
The energy-momentum tensor is given by the usual sense,
$T^{\mu\nu}=(2/\sqrt{-|g|}) \delta L_M/\delta g_{\mu\nu}.$

In this letter, we investigate the dynamics of a homogeneous scalar field in EiBI gravity.
The action for the matter field is then given by
\be \label{S:phi}
S_M = \int d^4 x \sqrt{-|g|}
\left[ -\frac12 g_{\mu\nu} \partial^\mu\phi \partial^\nu \phi -V(\phi) \right],
\ee
where the scalar field depends only on time, $\phi(t)$, because of the spatial homogeneity.
The homogeneous and isotropic ans\"atze for the metric and auxiliary metric are
\begin{align}
g_{\mu\nu} dx^\mu dx^\nu &= -dt^2 + a^2(t)\; d{\bf x}^2, \nn \\
q_{\mu\nu} dx^\mu dx^\nu &= -X^2(t)\; dt^2 + Y^2(t)\;  d{\bf x}^2.
\end{align}
From the EOM of 1st kind \eqref{eom1}, we obtain the auxiliary metric in terms of physical parameters,
\be\label{XY}
X = (\lambda-\kappa p)^{3/4}(\lambda+\kappa \rho)^{-1/4},\qquad
Y = \left[ (\lambda+\kappa \rho)(\lambda-\kappa p)\right]^{1/4} a,
\ee
where $\rho =\dot\phi^2/2 +V$ and $p = \dot\phi^2/2 -V$.
From the components of the EOM of 2nd kind~\eqref{eom2}, we obtain
the Hubble parameter,
\begin{eqnarray}\label{H}
H & \equiv& \frac{\dot a}{a} = \frac1{\sqrt{3\kappa}}
	\frac{1}{\left(\lambda/\kappa + V \right)^2 + \dot\phi^4/2}
\left\{ -\frac{\sqrt{3\kappa} }{2}\left(\frac{\lambda}{\kappa}+V + \frac{\dot \phi^2}{2}\right)
     V'(\phi)\dot \phi  \right.  \nn\\
     && \left. \pm \left(\frac{\lambda}{\kappa}+V-\frac{\dot \phi^2}2\right)
\Big[\kappa\left(\frac{\lambda}{\kappa}+V + \frac12\dot \phi^2\right)^{3/2}
\left(\frac{\lambda}{\kappa}+V-\frac{\dot \phi^2}2\right)^{3/2}
-\left(\frac{\lambda}{\kappa}+V + \frac12\dot \phi^2\right)
    \left( \frac{\lambda}{\kappa}+V -\dot \phi^2 \right)
     \Big]^{1/2}  \right\}.
\end{eqnarray}
This equation was first obtained in Eq.~(46) in Ref.~\cite{Scargill:2012kg} with $\lambda=1$.
There, it was also shown that GR is recovered in the leading order at later times of the expanding Universe.
Varying the action~\eqref{S:phi} with respect to $\phi$, the equation of motion for the scalar field is given by
\begin{equation} \label{ddphi:H}
\ddot \phi +3 H \dot \phi + V'(\phi) = 0.
\end{equation}

Now let us investigate the zero-curvature solution, i.e.,
the flat solution in terms of the metric $g_{\mu\nu}$.
We consider the flat spacetime, $a(t)={\rm constant}$,
with the scalar field $\phi(t)$ being dynamical.
This corresponds to $H=0$.
The equation~\eqref{H} allows a nontrivial solution for this, contrary to the case of GR in which $H=0$ at all times directly implies $\rho=0=p$.
The scalar-field equation \eqref{ddphi:H} can be integrated,
\be\label{dotphiSQ}
\ddot \phi = - V'(\phi)
\qquad\Rightarrow\qquad
\frac{1}{2}\dot \phi^2 = E- V(\phi),
\ee
where $E$ is the integration constant.
If we introduce an effective potential
$\cV \equiv (\lambda-\kappa E+ 2\kappa V)/c^2$,
where $c^2 \equiv \lambda +\kappa E$,
$H=0$ reduces to a simple form,
\begin{equation} \label{DV}
\frac{d\cV}{d\phi} = \pm
\sqrt{\frac{8}{3}}\;
	\frac{\cV \sqrt{1 -3\cV+2c^2{\cV}^{3/2}}}{c\sqrt{1 -\cV}}.
\end{equation}
The equation~\eqref{dotphiSQ} can be recast into
\be\label{KcV}
\frac{\kappa}{c^2} \dot\phi^2  +\cV(\phi) =1.
\ee
The rescaled dynamical field $(\sqrt{\kappa}/c) \phi$
with a fixed energy scale ${\cal E} =1$
subject to the potential $\cV(\phi)$ which is a solution to Eq.~\eqref{DV},
produces a {\it flat spacetime}.\footnote{From
Eq.~\eqref{KcV}, one gets
$\dot \phi = \pm (c/\sqrt{\kappa}) \sqrt{1-\cV}$.
Without loss of generality, one can take $c>0$.
Here, the signature change $+\to -$ is  equivalent to
the transformation $V(\phi) \to V(-\phi)$ accompanying $\phi\to -\phi$.
Therefore, in this work, we consider only the positive signature
which describes the positive velocity, $\dot \phi >0$.}

Let us comment on the values of $c=\sqrt{\lambda +\kappa E}$.
We shall consider the case of $\kappa >0$ in this work, and
assume that the energy density is non-negative, $\rho= E \geq 0$.
The cosmological constant becomes $\Lambda \gtreqqless 0$ for $\lambda \gtreqqless 1$.
Then one can have $c > 1$ for all types of the cosmological constant by tuning the energy density $\rho=E$.
One can have $c < 1$ only for the negative cosmological constant.
One can have $c=1$ for the negative cosmological constant with $E>0$,
or for the zero cosmological constant with $E=0$.

The shape of $d\cV /d\phi$ for $c>0$ is qualitatively different from that for $c \leq 1$.
The domain of $\cV$ is $[0,1]$.
For all the values of $c$, $d\cV /d\phi = 0$ at $\cV =0$.
For $c >1$, $d\cV /d\phi$  diverges to infinity at $\cV =1$.
For $c \leq 1$, $d\cV /d\phi$ possesses another zero
at $\cV_c$ where $1 -3\cV_c+2c^2{\cV_c}^{3/2}=0$.
[See Fig.~1(a).]

\begin{figure}
\begin{center}
\begin{tabular}{lr}
\includegraphics[width=.4\linewidth,origin=tl]{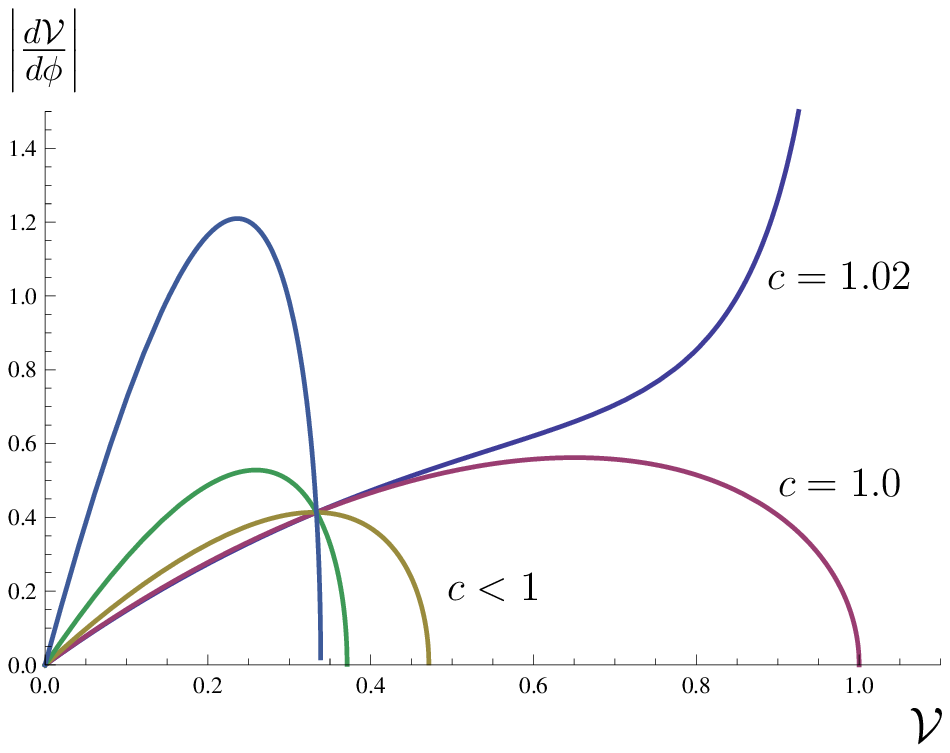} \qquad
&\includegraphics[width=.4\linewidth,origin=br]{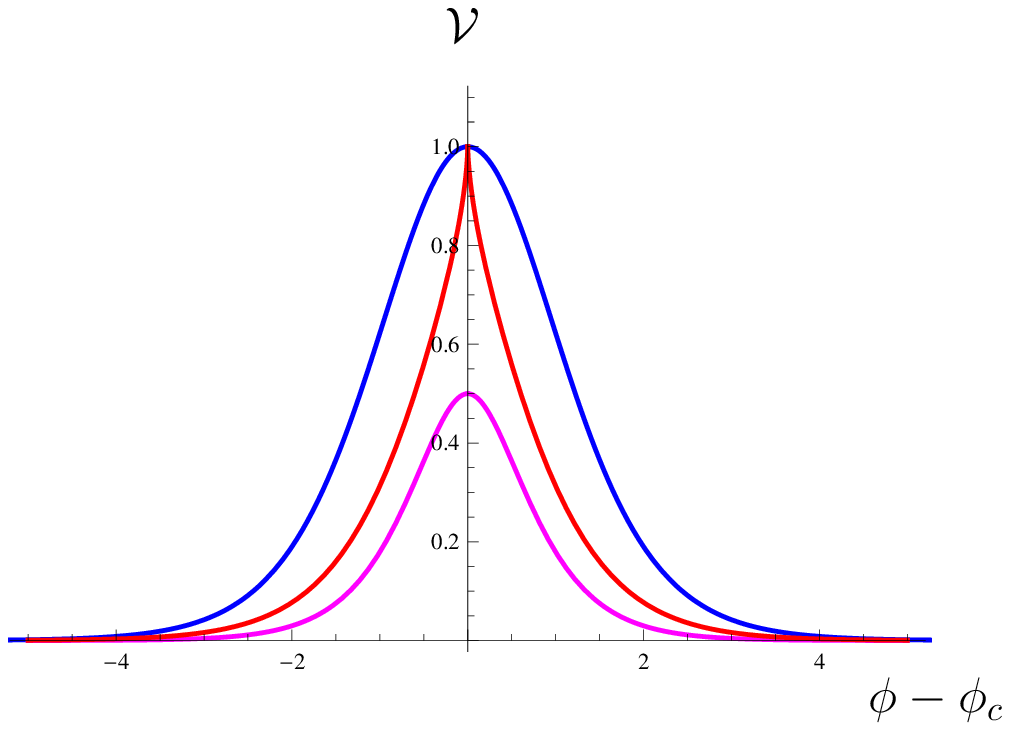}
\end{tabular}
\put(-330,-80) {$(a)$}
\put(-110,-80) {$(b)$}
\end{center}
\caption{
(a) Plot of $|d\mathcal{V}/d\phi|$ vs. $\cV$
for $c=1.02, 1.0, 0.8, 0.5, 0.2$.
(b) Plot of $\cV$ vs. $\phi$ for $c=1, 1.02, 0.84$, respectively from the top.
For the motion of the field $\phi$ subject to $\cV$,
the energy level is fixed to ${\cal E} =1$.
}
\label{Fig1}
\end{figure}

Now let us analyze the field dynamics and the effective potential from Eq.~\eqref{DV}.

(i)
For $c=1$, Eq.~\eqref{DV} is integrated in terms of Elliptic functions,
\be
2i \left[F(\arcsin(\sqrt{\frac{1+\cV^{1/2}}{2}}) |4)-\Pi(2;\arcsin(\sqrt{\frac{1+\cV^{1/2}}{2}})|4)-
F(\frac{\pi}{2}|4)+\Pi(2;\frac{\pi}{2}|4)\right]
	 =\pm \sqrt{\frac{2}{3}}\,(\phi -\phi_c),
\ee
where $F$ and $\Pi$ are the elliptic integral of first kind and the incomplete elliptic integral.
The solution $\cV(\phi)$ is plotted in Fig.1(b).
The field $\phi$ evolves in the positive direction with the energy level ${\cal E}=1$,
while the spacetime remains flat.
The field velocity becomes zero at $\phi=\phi_c$ and it takes infinite time in arriving there.
This point is an unstable extremum.

(ii) Near $\cV=0$ for all values of $c$, we have
\be\label{V1}
\frac{d\cV}{d\phi} \approx \pm
    \sqrt{\frac{8}{3}} \frac{\cV}{c}
\qquad\Rightarrow\qquad
\cV \approx \cV_0 e^{\pm \sqrt{\frac{8}3}\frac{\phi}{c}}.
\ee
Note that this solution is valid for $\cV \approx 0$, so for $|\phi| \gg c$.
The positive/negative signature corresponds to the left/right side of  $\cV$ in Fig.~1(b).
From $\dot \phi = (c/\sqrt{\kappa}) \sqrt{1-\cV}$, the scalar field becomes
\be\label{phi1}
\phi(t) \approx \frac{c}{\sqrt{\kappa}}t \mp
\sqrt{\frac{3}{32}} c\cV_0
e^{\pm \sqrt{\frac{8}{3\kappa}} t} \quad
    \mbox{ for } \quad  t \to \mp \infty.
\ee
The scalar field rolls up the exponential potential \eqref{V1}
for $t\ll - \sqrt{\kappa}/c$,
and rolls down for $t\gg  \sqrt{\kappa}/c$.

(iii)
Near $\cV =\cV_c \leq 1$ for $c \leq 1$, we have
\be\label{V2}
\frac{d\cV}{d\phi} \approx \pm \frac{\sqrt{\cV_c-\cV}}{c_1}
\qquad\Rightarrow\qquad
\cV \approx \cV_c -\frac{(\phi-\phi_c)^2}{4{c_1}^2}
\quad\mbox{ with }\quad
   \pm\phi \leq \pm\phi_c,
\ee
where $c_1=(c/\cV_c) [(1-\cV_c)/8/(1+c^2 \cV_c^{1/2} )]^{1/2}$.
The scalar field becomes then
\be
\phi(t) \approx \phi_c
+2c_1\sqrt{1-\cV_c}\, \sinh \left[\frac{c(t-t_c)}{2c_1 \sqrt{\kappa}} \right].
\ee
The top of the potential $\cV_c$ is reached at $\phi (t=t_c) =\phi_c$.
The field passes this point with nonvanishing velocity
except for $c=1$.

(iv)
Near $\cV =1$ for  $c >1$, we have
\be\label{V3}
\frac{\partial \cV}{\partial \phi} \approx \pm \frac1{c_2\sqrt{1-\cV}} ,
\qquad\Rightarrow\qquad
\cV \approx 1 - \left[\frac{3(\phi-\phi_c)}{2 c_2}\right]^{2/3}
\quad\mbox{ with }\quad
\pm\phi \leq \pm\phi_c,
\ee
where $c_2=\sqrt{3}c/(4\sqrt{c^2-1}) $.
The scalar field evolves as
\be
\phi \approx \phi_c \mp\frac{2}{3\sqrt{c_2}}
\left(\frac{c}{\sqrt{\kappa}} |t-t_c|\right)^{3/2} .
\ee
At the top of the potential, $\cV [\phi(t=t_c)=\phi_c]=1$,
the field velocity becomes zero ($\dot\phi =0$)
and the acceleration becomes infinite ($\ddot\phi = -d\cV/d\phi = \mp\infty$).
This point is unstable.

The dynamical scalar-field solutions that we obtained here
produces the flat spacetime.
Since EiBI gravity is equivalent to GR in vacuum,
the flat spacetime is also achieved in vacuum without a scalar field.
It means that there exist two branches of the flat spacetime.
Therefore, it is worthwhile to check the stability of the flat spacetime
originated from the dynamical scalar field.

In order to investigate the linear perturbation,
we introduce the linear perturbations $h(t)$ and $\psi(t)$
for the velocities of the metric and the scalar field,
\be
H(t) = 0 +\epsilon h(t),
\qquad
\dot \phi(t)=  \frac{c}{\sqrt{\kappa}} \sqrt{1- \cV}\; \left[1+ \epsilon\psi(t)\right],
\ee
and consider the field equations in the linear order in $\epsilon$.
The scalar-field equation~\eqref{ddphi:H} in the linear order is recast as
\be\label{psieq}
\frac{\dot \psi}{\psi} + \frac{3h}{\psi} -
	\frac{\dot\cV}{1-\cV}  =0.
\ee
The Hubble parameter \eqref{H} in the linear order is given by
\be\label{Hpsi}
\frac{h}{\psi} = \frac{\dot\cV}{3(1+\cV^2)-2 \cV  }
\left[ \frac{(1-\cV)(5-3\cV - 3c^2\cV^{1/2} (1-\cV))}{2(1-3 \cV+ 2c^2\cV^{3/2})}+\cV -1-\frac1{\cV}\right] .
\ee
Then Eq.~\eqref{psieq} can be integrated as
\begin{align}\label{psi}
\psi
&=\psi_0 \; \frac{\cV (1-3 \cV+ 2c^2\cV^{3/2})^{1/2}}{(1-\cV)(3-2\cV+3\cV^2)}.
\end{align}
\begin{figure}
\begin{center}
\begin{tabular}{ll}
\includegraphics[width=.4\linewidth,origin=tl]{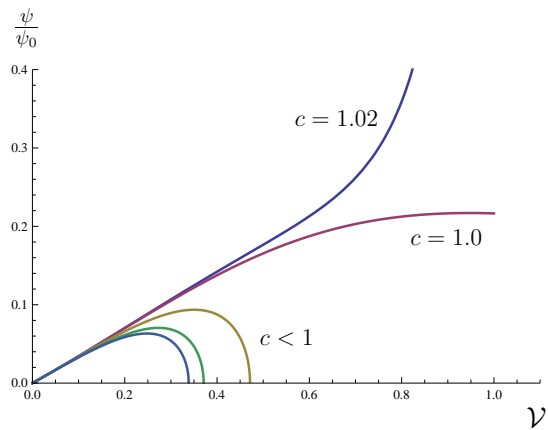}
\end{tabular}
\end{center}
\caption{The perturbation $\psi(t)$ for the same values of $c$ in Fig.~1(a).
The perturbation vanishes at $\cV=0$ and $\cV_c(<1)$.
}
\label{Fig2}
\end{figure}
The solution is plotted in Fig.~2.
The perturbation of the gravitational-field velocity $h(t)$ in Eq.~\eqref{Hpsi} becomes
\be\label{h}
h=\psi_0 \frac{-2 + 9 {\cal V} - 7 c^2 {\cal V}^{3/2} + 2 c^2 {\cal V}^{5/2} - 3 {\cal V}^3 + c^2 {\cal V}^{7/2}}{
2(1 - {\cal V}) (1 - 3 {\cal V} + 2 c^2 {\cal V}^{3/2})^{1/2} (3 - 2 {\cal V} + 3 {\cal V}^2)^2}
\;\dot\phi \; \frac{d\cV}{d\phi}.
\ee
The stability requires $\psi = h =0$.
From Eq.~\eqref{psi}, $\psi=0$ when $\cV=0$ and $\cV_c(<1)$.
From Eq.~\eqref{h}, $h=0$ when $\cV=\cV_*$ (the value of $\cV$ which makes the numerator zero)
and $\dot\cV = \dot\phi (d\cV/d\phi) =0$.
One can easily check that the system is complementarily {\it unstable};

(i) When $\cV=0$, $h\neq 0$.

(ii) When $\cV=\cV_c(<1)$, the scalar field $\phi$ just passes this point (the top of $\cV$) with a finite velocity.

(iii) When $\cV=\cV_*$, $\psi\neq 0$.

(iv) $\dot\cV = \dot\phi (d\cV/d\phi) =0$ is achieved at the top of $\cV$ for $c\leq 1$.
For $c<1$, it is unstable because of (ii). For $c=1$, it is unstable because $\psi \neq 0$.

\noindent
As a whole, the flat spacetime produced by the dynamical scalar field is unstable.

In summary, we have investigated the possibility that the nonlinearity of the matter-gravity interactions allows nongravitating matter distributions (nGMD).
We showed that nGMD is not possible in gravity theories such as general relativity, $f(R)$ gravity, and higher-derivative gravity, if they are regular in the zero-curvature limit.
On the other hand, the Eddington-inspired Born-Infeld gravity allows nGMD.
Explicitly, we found that a flat zero-curvature spacetime exists with nontrivial
scalar-field configurations subject to the potential $V(\phi)$ which satisfies the differential equation \eqref{DV}.
This potential is a runaway type as in Fig.~\ref{Fig1}(b).
For this configuration, the metric $g_{\mu\nu}$ is flat while the auxiliary metric $q_{\mu\nu}$ is nontrivial as Eq.~\eqref{XY}.
The spacetime curvature vanishes while the Ricci tensor evaluated by $q_{\mu\nu}$
is not trivial.
The modification of GR through the connection term by the Palatini formulation makes the EiBI theory allow nGMD.
The general gravity theories will be divided into two classes; the collections of gravity theories with and without nGMD.
The EiBI gravity will be the first example belonging to the class with nGMD.

Since EiBI gravity is equivalent to GR in vacuum, the flat spacetime is also achieved when there is {\it no matter}.
Therefore, there are two branches of the flat spacetime.
The nGMD solution obtained in this work is unstable under perturbations.
This unstable one will evolve to other state.
It would be interesting to study how this evolution proceed and what the final state will be.
Another interesting question is whether or not the EiBI gravity allows a nongravitating localized object such as a boson star.

\subsection*{Acknowledgements}
This work was supported by the Korea Research Foundation (KRF) grants
funded by the Korea government (MEST) No. 2012-006136 (I.C.) and
No. 2010-0011308 (H.K.)


\begin{thebibliography}{99}

\bibitem{Banados:2010ix}
  M.~Banados and P.~G.~Ferreira,
  Phys.\ Rev.\ Lett.\  {\bf 105}, 011101 (2010)
  [arXiv:1006.1769 [astro-ph.CO]].

\bibitem{Cho:2012vg}
  I.~Cho, H.~-C.~Kim and T.~Moon,
  Phys.\ Rev.\ D {\bf 86}, 084018 (2012)
  [arXiv:1208.2146 [gr-qc]].

\bibitem{DeFelice:2012hq}
  A.~De Felice, B.~Gumjudpai and S.~Jhingan,
  Phys.\ Rev.\ D {\bf 86}, 043525 (2012)  [arXiv:1205.1168 [gr-qc]].  

\bibitem{Avelino:2012ge}
  P.~P.~Avelino,
  Phys.\ Rev.\ D {\bf 85}, 104053 (2012)  [arXiv:1201.2544 [astro-ph.CO]];
  P.~P.~Avelino,
  JCAP {\bf 1211}, 022 (2012)  [arXiv:1207.4730 [astro-ph.CO]].

\bibitem{Casanellas:2011kf}
  J.~Casanellas, P.~Pani, I.~Lopes and V.~Cardoso,
  Astrophys.\ J.\  {\bf 745}, 15 (2012)
  [arXiv:1109.0249 [astro-ph.SR]].


\bibitem{EscamillaRivera:2012vz}
  C.~Escamilla-Rivera, M.~Banados and P.~G.~Ferreira,
  Phys.\ Rev.\ D {\bf 85}, 087302 (2012)
  [arXiv:1204.1691 [gr-qc]].


\bibitem{Avelino:2012ue}
  P.~P.~Avelino and R.~Z.~Ferreira,
  Phys.\ Rev.\ D {\bf 86}, 041501 (2012)  [arXiv:1205.6676 [astro-ph.CO]].


\bibitem{Liu:2012rc}
  Y.~-X.~Liu, K.~Yang, H.~Guo and Y.~Zhong,
  Phys.\ Rev.\ D {\bf 85}, 124053 (2012)
  [arXiv:1203.2349 [hep-th]].

\bibitem{Delsate:2012ky}
  T.~Delsate and J.~Steinhoff,
  Phys.\ Rev.\ Lett.\  {\bf 109}, 021101 (2012)  [arXiv:1201.4989 [gr-qc]].

\bibitem{Scargill:2012kg}
  J.~H.~C.~Scargill, M.~Banados and P.~G.~Ferreira,
  Phys.\ Rev.\ D {\bf 86}, 103533 (2012)  [arXiv:1210.1521 [astro-ph.CO]].

\bibitem{Sham:2012qi}
  Y.~-H.~Sham, L.~-M.~Lin and P.~T.~Leung,
  Phys.\ Rev.\ D {\bf 86}, 064015 (2012)  [arXiv:1208.1314 [gr-qc]].

\bibitem{Pani:2012qd}
  P.~Pani and T.~P.~Sotiriou,
  Phys.\ Rev.\ Lett.\  {\bf 109}, 251102 (2012)  [arXiv:1209.2972 [gr-qc]].


\end{thebibliography}
\end{document}